\documentclass[11pt,a4paper,notitlepage]{article}

\usepackage[activeacute,english]{babel}

\usepackage[latin1]{inputenc}
\usepackage[active]{srcltx}
\usepackage[toc,page]{appendix}
\usepackage{fancyhdr}
\usepackage{amsmath}
\usepackage{amsfonts}
\usepackage{amssymb}
\usepackage{graphicx}
\usepackage{caption}
\usepackage{subcaption}
\usepackage{xcolor}

\usepackage{skak}
\usepackage{MnSymbol}
\usepackage{bbding}
\usepackage{skull}

\usepackage[normalem]{ulem}
\usepackage{authblk}
\makeatletter
\let\old@startsection=\@startsection
\let\oldl@section=\l@section
\renewcommand{\@startsection}[6]{\old@startsection{#1}{#2}{#3}{#4}{#5}{#6\mathversion{bold}}}
\renewcommand{\l@section}[2]{\oldl@section{\mathversion{bold}#1}{#2}}
\makeatother

\makeatletter \@addtoreset{equation}{section} \makeatother

\title{\textbf{Quantum Information Metric and Berry Curvature from a Lagrangian Approach}}

\author{Javier Alvarez-Jimenez$^{\,1}$, Aldo Dector$^{\,2}$, J. David Vergara$^{\,1}$ \thanks{Email address: aldo.dector@gmail.com, vergara@nucleares.unam.mx }}

\affil{\textit{1) Instituto de Ciencias Nucleares\\
\textit{Universidad Nacional Autonoma de M\'{e}xico}\\
\textit{Ciudad de M\'{e}xico, M\'{e}xico}}}

\affil{\textit{2) Instituto de F\'{i}sica Te\'{o}rica IFT UAM/CSIC\\
\textit{Calle Nicol\'{a}s Cabrera 13. UAM, Cantoblanco 28049}\\
\textit{ Madrid, Spain}}}

\date{}

\begin{document}
\begin{titlepage}
    \maketitle

    \begin{abstract}
		
We take as a starting point an expression for the quantum geometric tensor recently derived in the context of the gauge/gravity duality. We proceed to generalize this formalism in such way it is  possible to compute the geometrical phases of quantum systems. Our scheme provides a conceptually complete description and introduces a different point of view of earlier works. Using our formalism, we show how this expression can be applied to well-known quantum mechanical systems. 
		
    \end{abstract}
  \end{titlepage}

\clearpage

\tableofcontents


\section{Introduction}

Understanding how information is encoded in a quantum system is of fundamental importance in quantum mechanics and quantum field theory. Two essential elements to extract this information are the quantum information metric \cite{Provost}  and the geometric phases \cite{Berry, Simon, Anandan, Wilczek}. Both tools have been very useful to analyze classical \cite{Hannay, Benenti}  and quantum systems \cite{Gu}.  For example, at the quantum level the quantum information metric is an important tool to explore quantum phase transitions \cite{Zanardi, Gu, Sachdev, Kumar:2012ve, Maity:2015rfa} and the geometric phases control a key effect in quantum mechanics, the quantum interference \cite{Chru}. On the other hand it is interesting to observe that within the approach of Provost and Vallee \cite{Provost} both structures are part of the same geometrical structure, a complex quantum geometric tensor, whose real part corresponds to the quantum information metric, while its imaginary part corresponds to the Berry curvature. However, in the usual treatments both entities are calculated differently, see for example \cite{Moore} and \cite{Gu}. In this article, within a Lagrangian approach recently introduced in \cite{MIyaji:2015mia} and further developed in \cite{Bak:2015jxd, Trivella:2016brw}, we consider how to give a unified treatment of the quantum information metric and the Berry curvature. We start from the approach developed in \cite{MIyaji:2015mia, Bak:2015jxd, Trivella:2016brw} to the computation of the Quantum Information Metric (QIM) and we generalize this formalism to include the Berry curvature and show using several quantum mechanical examples that our formalism produces the correct results. Our scheme provides a conceptually complete description and introduces a different point of view of earlier works, see for instance \cite{Seligman}.  Our proposal is in some sense related to the one given in \cite{Kolodrubetz}; nevertheless, our approach uses essentially the path integral in the Lagrangian form. Also in the context of the path-integral there have been some attempts to compute the quantum fidelity amplitude \cite{Vanicek}, but we consider that our approach is more natural and includes not only the fidelity but also the full quantum information metric tensor and the geometric phases. We apply our formalism to study the quantum metric tensor and the geometric phases in different quantum mechanical systems modified by well-defined deformation terms. The main point would be to study a quantum mechanical system presenting a quantum phase transition. For more on possible systems, see \cite{Gu}.

The structure of this paper is the following. In Section 2, we show the derivation given in  \cite{MIyaji:2015mia, Bak:2015jxd, Trivella:2016brw}, and in the subsection 2.2 we introduce our generalization, so that in the same scheme we can compute 
the quantum information metric and the Berry curvature. In Section 3, we show explicitly how our formalism works in the case of the quantum harmonic oscillator with a linear deformation.  Section 4 is devoted to study the generalized harmonic oscillator, where we compute the quantum information metric and the Berry curvature. In Section 5, we consider the quantum XY model in a transverse magnetic field and we compute its quantum information metric. In Section 6, we analyze the Ising model in a transverse field adiabatically rotated and we calculate the Berry curvature. Finally in Section 7, we present our conclusions.


\section{A Lagrangian Approach to Quantum Information Metric and the Berry Curvature}

\subsection{Review of the Original Derivation: General Results for $D=d+1$ Quantum Field Theories}

In this section we review a Lagrangian approach to the quantum fidelity amplitude, as originally proposed in \cite{MIyaji:2015mia} and further studied in \cite{Bak:2015jxd, Trivella:2016brw, Banerjee:2017qti} within the context of the gauge/gravity duality. We will start by considering the very general case of a quantum field theory in $D=d+1$ dimensions in Euclidean time as in the original derivation, and later we will focus on the particular case of a $D=0+1$ quantum mechanical system. For an alternate path-integral approach to the quantum fidelity, see \cite{Vanicek}. For more research on geometric phases and the information metric from the point of view of quantum field theory, see \cite{Balasubramanian:2014bfa, Baggio:2017aww}.

The original physical setup is the following. We start by considering a quantum field theory defined by a Lagrangian $\mathcal{L}_{0}$ during the Euclidean time interval $(-\infty,0)$. We can then consider the situation when a deformation of the original theory is turned on at time $\tau=0$. This deformation is enacted by the addition to the initial Lagrangian $\mathcal{L}_{0}$ of terms of the form $\delta\lambda^{a} \mathcal{O}_{a}$, with parameter-space index $a=1,\,\ldots\, n$, where $\mathcal{O}_{a}$ are deformation operators which are functions of the theory's degrees of freedom, and $\delta\lambda^{a}$ are real parameters associated with these deformations. In the context of the systems considered in quantum information theory we should consider the situation where the original Lagrangian $\mathcal{L}_{0}$ has an explicit dependence on a number of physical parameters $\lambda^{a}$ like frequency, magnetic field, etc., so that the final deformed theory arises from small variations of these parameters by $\lambda^{a}\rightarrow\lambda^{a}+\delta \lambda^{a}$. In this manner, we obtain a perturbed Lagrangian $\mathcal{L}_{1}$ for the remaining Euclidean time interval $(0,\infty)$, given explicitly by
\begin{equation} 
\label{deformedL}
\mathcal{L}_{1}=\mathcal{L}_{0}+\delta \lambda^{a}\, \mathcal{O}_{a}\,.
\end{equation}

Both the original and deformed theories have corresponding states $|\,\Omega_{0}\rangle$ and $|\,\Omega_{1}\rangle$, respectively. We will now focus on computing the fidelity of this system, defined here as the absolute square of the overlap between both states at temporal infinity $\mathcal{F}(\lambda,\,\lambda+\delta\lambda)\equiv|\langle \Omega_{1},\tau\rightarrow\infty|\Omega_{0},\tau\rightarrow -\infty\rangle|$. Physically speaking, the fidelity thus defined will give us a measure of the change effectuated on the system by turning on the deformation terms. Following the original derivation, we will begin by considering any generic state $|\varphi\rangle$ and its overlap with the original state $|\,\Omega_{0}\rangle$. In the path-integral formalism, this can be obtained by evolving the system in Euclidean time from $\tau=-\infty$ where the original state is placed, to $\tau=0$ where we take the state $|\varphi\rangle$ to be inserted. This is written as
\begin{equation}
\langle\varphi \,|\, \Omega_{0}\rangle = \frac{1}{\sqrt{Z_{0}}}\int_{\varphi(\tau = 0)=\tilde{\varphi}}\mathcal{D}\varphi \exp\left(-\int_{-\infty}^{0}d\tau \int d^{d}x\; \mathcal{L}_{0}\right)\,,
\end{equation}
where
\begin{equation}
Z_{0}=\int \mathcal{D}\varphi\exp \left(-\int _{-\infty}^{\infty} d\tau \int d^{d}x\;\mathcal{L}_{0}\right)\,.
\end{equation}
Similarly, one can consider the evolution from $\tau=0$ where the state $|\varphi\rangle$ is inserted, to $\tau=\infty$ where we are placing the perturbed state $|\,\Omega_{1}\rangle$. In the path integral-formalism this is given by
\begin{eqnarray}
\langle\,\Omega_{1}|\, \varphi\rangle &=& \frac{1}{\sqrt{Z_{1}}}\int_{\varphi(\tau = 0)=\tilde{\varphi}}\mathcal{D}\varphi \exp\left(-\int_{0}^{\infty}d\tau \int d^{d}x\; \mathcal{L}_{1}\right)\nonumber\\
&=& \frac{1}{\sqrt{Z_{1}}}\int_{\varphi(\tau = 0)=\tilde{\varphi}}\mathcal{D}\varphi \exp\left(-\int_{0}^{\infty}d\tau \int d^{d}x\; \left(\mathcal{L}_{0}+\delta \lambda^{a}\mathcal{O}_{a}\right)\right)\,,
\end{eqnarray}
where
\begin{equation}
Z_{1}=\int \mathcal{D}\exp \left(-\int _{-\infty}^{\infty} d\tau \int d^{d}x\; \left(\mathcal{L}_{0}+\delta \lambda^{a}\mathcal{O}_{a}\right)\right)\,.
\end{equation}
Thus, the overlap between both states is given by
\begin{eqnarray}
\langle \Omega_{1}|\Omega_{0}\rangle &=& \int \mathcal{D}\varphi \;\langle \Omega_{1}|\varphi\rangle\langle\varphi|\Omega_{0}\rangle\nonumber\\
&=&\frac{1}{\sqrt{Z_{0}Z_{1}}}\int \mathcal{D}\varphi \;\exp \left(-\int_{-\infty}^{0}d\tau\int d^{d}x\,\mathcal{L}_{0}-\int_{0}^{\infty}d\tau \int d^{d}x \left(\mathcal{L}_{0}+\delta\lambda^{a}\mathcal{O}_{a}\right) \right)\,.\nonumber\\
\end{eqnarray}
which can be written more conveniently as
\begin{equation}
\label{overlap}
\langle \Omega_{1}|\Omega_{0}\rangle=\frac{\big\langle \exp\left(-\int_{0}^{\infty}d\tau \int d^{d}x\; \delta\lambda^{a}\mathcal{O}_{a}\right)\big\rangle}{\big\langle \exp \left(-\int_{-\infty}^{\infty}d\tau \int d^{d}x\, \delta\lambda^{a}\mathcal{O}_{a}\right)\big\rangle^{1/2}}\,,
\end{equation}
where the expectation value is taken with respect to the unperturbed state $|\Omega_{0}\rangle$.\footnote{It is important to note that there is a discontinuity when passing from the original to the deformed Lagrangian at Euclidean time $\tau=0$, when we suddenly turn on the deformation, which could result in ultraviolet divergences. The formal way to address this is to introduce a cutoff scale $\epsilon$ around $\tau=0$, removing the region where the Lagrangian changes suddenly. The introduction of this cutoff scale $\epsilon$ is of great importance in the gauge/gravity duality context of the original derivation. However, for simplicity, in the following we will assume that this scale is indeed formally present, but taken to be zero when seen in comparison with the natural scales of each system under consideration.}

We can now take expression (\ref{overlap}) to expand $|\langle \Omega_{1}|\Omega_{0}\rangle|$ in a series of $\delta\lambda^{a}$. In the original derivation, where the authors studied the particular case of conformal field theories such as the ones considered in the gauge/gravity correspondence, one has $\langle \mathcal{O}_{a}\rangle=0$ for an operator of non-zero dimension. However, in the following we will consider the more general case where the expectation values can be different from zero.  Also, we will assume that the two-point functions have time-reversal symmetry $\langle \mathcal{O}_{a}(-\tau_{1})\mathcal{O}_{b}(-\tau_{2})\rangle=\langle \mathcal{O}_{a}(\tau_{1})\mathcal{O}_{b}(\tau_{2})\rangle$. 

The final result of this expansion of the fidelity can be written as
\begin{equation}
\label{QFTfid}
|\langle \Omega_{1}|\Omega_{0}\rangle|=1-\frac{1}{2}G_{a b}\,\delta\lambda^{a}\delta\lambda^{b}+\cdots\,.
\end{equation}
\bigskip
and where the quantity $G_{ab}$ is known as the \textit{complex quantum geometric tensor} and is given here by
\begin{equation}
\label{QFTQIM}
G_{a b}=\int_{-\infty}^{0}d\tau_{1}\int_{0}^{\infty}d\tau_{2}\int d^d x_1 \int d^d x_2 \,\Big(\langle \mathcal{O}_{a}(\tau_{1},x_{1})\mathcal{O}_{b}(\tau_{2},x_{2})\rangle-\langle\mathcal{O}_{a}(\tau_{1},x_{1})\rangle\langle\mathcal{O}_{b}(\tau_{2},x_{2})\rangle\Big)\,,
\end{equation}
where there is an implicit time ordering in the two-point function, given that this expression is obtained from a path-integral approach.

In the following subsection we will show how both the quantum information metric and the Berry curvature of a physical system can be calculated from this two-point function expression (\ref{QFTQIM}). For a brief review on the derivation these quantities, the reader is referred to the Appendix.


\subsection{Quantum Mechanical Systems: The Quantum Information Metric and the Berry Curvature}

We now consider the particular case of a $D=0+1$ quantum field theory, i.e. a quantum mechanical system. We should note that the passing from Minkowski to Euclidean time in the quantum field theory context of the previous section should be referred to as performing a Wick rotation in the context of quantum mechanical systems. As a matter of notation, we will refer with $t$ to real time, and refer with $\tau$ to Wick-rotated time. Therefore, in the case of quantum mechanical systems the quantum geometric tensor (\ref{QFTQIM}) reduces to the form
\begin{equation}
\label{QMFS}
G_{a b}=\int_{-\infty}^{0}d\tau_{1}\int_{0}^{\infty}d\tau_{2}\,\Big(\langle \mathcal{O}_{a}(\tau_{1})\mathcal{O}_{b}(\tau_{2})\rangle-\langle \mathcal{O}_{a}(\tau) \rangle\langle \mathcal{O}_{b}(\tau) \rangle\Big)\,.
\end{equation}

In analogy with the standard formulation of the quantum geometric tensor (see Appendix for further details), we intend to divide the expression (\ref{QMFS}) in its real and imaginary parts. In doing this we will make implicit use of time reversal symmetry for the two-point functions and the fact that the deformation operators are Hermitian. The real part can be written as
\begin{eqnarray}
\text{\textbf{Re}}\, G_{a b}&=&\int_{-\infty}^{0}d\tau_{1}\int_{0}^{\infty}d\tau_{2}\,\left(\frac{1}{2}\Big(\langle\mathcal{O}_{a}(\tau_1)\mathcal{O}_{b}(\tau_2)\rangle+\langle\mathcal{O}_{b}(\tau_1)\mathcal{O}_{a}(\tau_2)\rangle\Big)-\langle \mathcal{O}_{a}(\tau_{1})\rangle\langle\mathcal{O}_{b}(\tau_{2})\rangle\right)\nonumber\\
\label{ReZI}
&=&\int_{-\infty}^{0}d\tau_{1}\int_{0}^{\infty}d\tau_{2}\,\left(\frac{1}{2}\langle\left\{\mathcal{O}_{a}(\tau_1),\mathcal{O}_{b}(\tau_2)\right\}\rangle-\langle \mathcal{O}_{a}(\tau_{1})\rangle\langle\mathcal{O}_{b}(\tau_{2})\rangle\right)\,,
\end{eqnarray}
whereas the imaginary part of $G_{ab}$ takes the form
\begin{eqnarray}
\text{\textbf{Im}}\,G_{a b}&=&\frac{1}{2i}\int_{-\infty}^{0}d\tau_{1}\int_{0}^{\infty}d\tau_{2}\,\Big(\langle\mathcal{O}_{a}(\tau_1)\mathcal{O}_{b}(\tau_2)\rangle-\langle\mathcal{O}_{b}(\tau_1)\mathcal{O}_{a}(\tau_2)\rangle\Big)\nonumber\\
\label{ImzI}
&=&\frac{1}{2i}\int_{-\infty}^{0}d\tau_{1}\int_{0}^{\infty}d\tau_{2}\,\langle\left[\mathcal{O}_{a}(\tau_1),\mathcal{O}_{b}(\tau_2)\right]\rangle\,,
\end{eqnarray}

We here note that the expression (\ref{ReZI}) for the imaginary part of $G_{ab}$ involves the anticommutator $\left\{\;,\;\right\}$ between the deformation operators, while expression (\ref{ImzI}) for the imaginary part involves the commutator $\left[\;,\;\right]$. We also should point to the fact that, since we are working in a formalism originally derived from a path-integral approach, the time ordering in the two-point functions involved in $	\textbf{Re}\,G_{ab}$ and $\textbf{Im} G_{ab}$ presented above should be the same as in the original expression for the quantum geometric tensor (\ref{QMFS}).

We see in this manner that $\text{\textbf{Re}}\,G_{ab}$ is symmetric on the parameter space indexes, in analogy with the expression (\ref{standardG}) for the quantum information metric of  a system and that $\text{\textbf{Im}}\,G_{ab}$ is antisymmetric in these same indexes, in analogy with the expression (\ref{standardF}) for the Berry curvature of the system. We then make the claim that the real part (\ref{ReZI}) corresponds to the \textit{quantum information metric}, $\textbf{Re}\,G_{ab}=g_{ab}$, that is
\begin{equation}
\label{ReZ}
g_{ab}=\int_{-\infty}^{0}d\tau_{1}\int_{0}^{\infty}d\tau_{2}\,\left(\frac{1}{2}\langle\left\{\mathcal{O}_{a}(\tau_1),\mathcal{O}_{b}(\tau_2)\right\}\rangle-\langle \mathcal{O}_{a}(\tau_{1})\rangle\langle\mathcal{O}_{b}(\tau_{2})\rangle\right)\,,
\end{equation}
while the imaginary part (\ref{ImzI}) corresponds to the \textit{Berry curvature}, $\textbf{Im}\,G_{ab}=\frac{1}{2}F_{ab}$, explicitly
\begin{equation}
\label{ImZ}
F_{ab}=\frac{1}{i}\int_{-\infty}^{0}d\tau_{1}\int_{0}^{\infty}d\tau_{2}\,\langle\left[\mathcal{O}_{a}(\tau_1),\mathcal{O}_{b}(\tau_2)\right]\rangle\,.
\end{equation}
Its is important to mention that, according to this prescription, it is necessary to have more that one parameter in order for the system to exhibit a geometric phase.

In the following sections we intent to prove this last assertion by computing with this method the quantum information metric and Berry curvature in various different quantum mechanical setups.


\section{The Quantum Harmonic Oscillator: Linear Deformation}

\subsection{General Setting and Solution}

We begin by studying a simple example of a deformation to the quantum harmonic oscillator. This will allow us to become familiarized with some of the basic techniques and procedures related to the application of the Lagrangian formalism. In order to set notation, we will write the undeformed Hamiltonian $H_{0}$ of the system as
\begin{equation}
\label{QHO}
H_{0}=\frac{1}{2}p^{2}+\frac{\alpha}{2}q^{2}\,,
\end{equation}
where we are working in units where $\hbar =1$ and have set any parameters different from the real number $\alpha$ in the Hamiltonian equal to unity. The simplest deformation we can introduce in Hamiltonian (\ref{QHO}) is a term which is linear in the position operator, so that the resulting deformed theory is described by
\begin{equation}
\label{HOlinear}
H_{1}=\frac{1}{2}p^{2}+\frac{\alpha}{2}q^{2}+J\, q\,,
\end{equation}
where $J$ is a real parameter.  Clearly, this is a case where the corresponding deformed Lagrangian can be written in the desired form (\ref{deformedL})
\begin{equation}
L_{1} = L_{0}+J \mathcal{O}_{J}\,,\hspace{20pt}\text{with}\hspace{20pt}\mathcal{O}_{J}=-q\,.
\end{equation}
We should mention that, given that the deformation has only one parameter, we then cannot expect to find a Berry curvature in this example. The Hamiltonian (\ref{HOlinear}) can be rewritten as
\begin{equation}
H_{1}=\frac{1}{2}p^{2}+\frac{\alpha}{2}Q^{2}-\frac{1}{2}\left(\frac{J}{\sqrt{\alpha}}\right)^{2}\,,
\end{equation}
where
\begin{equation}
Q\equiv q+\frac{J}{\alpha}\,.
\end{equation}
Let $|\Psi_{n}^{(0)}\rangle$ and $|\Psi_{m}^{(1)}\rangle$ be the eigenfunctions of the $H_{0}$ and $H_{1}$ operators, respectively. Then, these eigenfunctions satisfy 
\begin{equation}
H_{0}|\Psi_{n}^{(0)}\rangle=E^{(0)}_{n}|\Psi_{n}^{(0)}\rangle\,,\hspace{20pt}H_{1}|\Psi_{m}^{(1)}\rangle=E^{(1)}_{m}|\Psi_{m}^{(1)}\rangle\,,
\end{equation}
with corresponding eigenvalues
\begin{equation}
E_{n}^{(0)}=\sqrt{\alpha}\left(n+\frac{1}{2}\right)\,,\hspace{25pt}E_{m}^{(1)}=\sqrt{\alpha}\left(m+\frac{1}{2}\right)-\frac{1}{2}\left(\frac{J}{\sqrt{\alpha}}\right)^{2}\,.
\end{equation}

We can now focus on the ground-state functions of each Hamiltonians in the coordinate representation
\begin{eqnarray}
\Psi_{0}^{(0)}(q)&=&\left(\frac{\sqrt{\alpha}}{\pi}\right)^{1/4}\text{Exp}\left\{-\frac{\sqrt{\alpha}}{2}q^{2}\right\}\,,\\
\Psi_{0}^{(1)}(q)&=&\left(\frac{\sqrt{\alpha}}{\pi}\right)^{1/4}\text{Exp}\left\{-\frac{\sqrt{\alpha}}{2}\left(q+\frac{J}{\alpha}\right)^{2}\right\}\,.
\end{eqnarray}

With these exact solutions we can compute directly the overlap between the different ground states in order to compute the system's fidelity when the deformation is turned on.\footnote{The case of the overlap of the unperturbed and perturbed states at $\tau=-\infty$ and $\tau=+\infty$ respectively, differs from the case where both states are at $\tau=0$ by a phase that disappears when taking the absolute value.} In this manner, taking the absolute square of the overlap and expanding in series of the parameter $J$ we obtain
\begin{equation}
|\langle \Psi_{0}^{(1)}|\Psi_{0}^{(0)}\rangle|= 1-\frac{1}{4\alpha^{3/2}}J^{2}+\cdots\,,
\end{equation}
so that, comparing with expression (\ref{QFTfid}), we can directly read the quantum information metric $g_{JJ}$  associated with the single parameter $J$ as
\begin{equation}
\label{QIMResI}
g_{JJ}=\frac{1}{2\alpha^{3/2}}\,.
\end{equation}

\subsection{Quantum Information Metric from the Lagrangian Formalism}

We can now realize the computation of the quantum information metric using the expression (\ref{ReZ}) we have obtained from the Lagrangian formalism.
\begin{eqnarray}
g_{JJ}&=&\int_{-\infty}^{0}d\tau_{1}\int_{0}^{\infty}d\tau_{2}\left(\langle \Psi_{0}^{(0)}| \mathcal{O}_{J}(\tau_{1})\mathcal{O}_{J}(\tau_{2})|\Psi_{0}^{(0)}\rangle-\langle \Psi_{0}^{(0)}| \mathcal{O}_{J}(\tau_{1})| \Psi_{0}^{(0)}\rangle \langle \Psi_{0}^{(0)}| \mathcal{O}_{J}(\tau_{2})| \Psi_{0}^{(0)}\rangle\right)\nonumber\\
\label{LinearMetric}
&=& \int_{-\infty}^{0}d\tau_{1}\int_{0}^{\infty}d\tau_{2}\left(\langle \Psi_{0}^{(0)}| q(\tau_{1})q(\tau_{2})|\Psi_{0}^{(0)}\rangle-\langle \Psi_{0}^{(0)}|q(\tau_{1})|\Psi_{0}^{(0)}\rangle \langle \Psi_{0}^{(0)}| q(\tau_{2})| \Psi_{0}^{(0)}\rangle\right)\,,
\end{eqnarray}
where the expectation value is taken with respect to the undeformed ground state $|\Psi_{0}^{(0)}\rangle$. In order to compute (\ref{LinearMetric}), we first note that the two point function of the quantum harmonic oscillator with respect to its ground state in real time is given by
\begin{equation}
\label{QHOtwopoint}
\langle \Psi_{0}^{(0)} |\,q(t_{1})q(t_{2})\,|\Psi_{0}^{(0)}\rangle=\frac{1}{i}G_{2}(t_{2}-t_{1})\,
\end{equation}
where $G_{2}(t_{2}-t_{1})$ is the Green's function for the harmonic oscillator
\begin{equation}
\label{HOgreen}
G_{2}(t_{2}-t_{1})=\frac{i}{2\sqrt{\alpha}}\text{Exp}\left(-i\sqrt{\alpha}|t_{2}-t_{1}|\right)\,,
\end{equation}
so that making a Wick rotation $t = -i\tau $ we obtain
\begin{equation}
\label{QHOtwopointWick}
\langle \Psi_{0}^{(0)} |\,q(\tau_{1})q(\tau_{2})\,|\Psi_{0}^{(0)}\rangle=\frac{1}{2\sqrt{\alpha}}\text{Exp}\left(-\sqrt{\alpha}(\tau_{2}-\tau_{1})\right)\,,\hspace{20pt}\tau_{2}>\tau_{1}\,,
\end{equation}
while one can find $\langle \Psi_{0}^{(0)}|\,q\,| \Psi_{0}^{(0)}\rangle=0$. Therefore, substituting (\ref{QHOtwopointWick}) in expression (\ref{LinearMetric}) for the quantum information metric we obtain
\begin{eqnarray}
g_{JJ}&=&\int_{-\infty}^{0}d\tau_{1}\int_{0}^{\infty}d\tau_{2}\,\langle \Psi_{0}^{(0)} |\,q(\tau_{1})q(\tau_{2})\,|\Psi_{0}^{(0)}\rangle\nonumber\\
&=&\frac{1}{2\sqrt{\alpha}}\int_{-\infty}^{0}d\tau_{1}\; e^{\sqrt{\alpha} t_{1}}\int_{0}^{\infty}d\tau_{2}\; e^{-\sqrt{\alpha} t_{2}}\nonumber\\
&=&\frac{1}{2\alpha^{3/2}}\,,
\end{eqnarray}
which is exactly the same expression as (\ref{QIMResI}). Therefore, we can see in this example that the application of the Lagrangian formalism can give us the same result for the quantum information metric obtained through direct computation.


\section{The Generalized Harmonic Oscillator}

\subsection{General Setting and Solution}

We will now focus on the study of the more interesting system known as the \textit{generalized harmonic oscillator}. The interested reader can consult \cite{Chru} for further details. See also \cite{Morales}. The Hamiltonian of this system is given by
\begin{equation}
\label{GHO}
H= Z p^2 + Y \left\lbrace p,q \right\rbrace + X q^2\,,
\end{equation}
where $X,Y,Z$ are real numbers that act as the parameters of the system. To obtain the Lagrangian, we have from Hamilton's equations 
\begin{equation}
\label{momentum}
p=\frac{1}{2Z}\left(\dot{q}-2Y q\right)\,,
\end{equation}
so that substituting in $L(q,\dot{q})=\dot{q}p\left(q,\dot{q}\right)-H\left(q,\dot{q}\right)$ we can write
\begin{equation}
L=\frac{1}{4Z}\dot{q}^{2}-\left(X-\frac{Y^{2}}{Z}\right)q^{2}-\frac{Y}{Z}\dot{q}q\,.
\end{equation}

We can now do small variations of the parameters $\{X,Y,Z\}\rightarrow\{X+\delta X,Y+\delta Y,Z+\delta Z\}$ and keep terms up to linear order. Doing this and writing the deformed Lagrangian in the form (\ref{deformedL}), we can read the deformation operators $\mathcal{O}_{a}$ as
\begin{eqnarray}
\mathcal{O}_{X}&=&-q^{2}\,,\\
\mathcal{O}_{Y}&=&-\frac{q}{2}\left(\dot{q}-2Y q\right)\,,\\
\mathcal{O}_{Z}&=&-\frac{1}{4Z^{2}}\left(\dot{q}-2Y q\right)^{2}\,.
\end{eqnarray}
Note that these operators can be written in term of the momentum (\ref{momentum}) as:
\begin{eqnarray}
\mathcal{O}_{X}&=&-q^{2}\,,\label{Ox}\\
\mathcal{O}_{Y}&=&-2 q p = - \left\{q,p\right\}\,,\label{Oy}\\
\mathcal{O}_{Z}&=&-p^{2}\,,\label{Oz}
\end{eqnarray}
as should be expected from inspection of the original Hamiltonian (\ref{GHO}). 

The generalized harmonic oscillator can be solved exactly, and below we will very briefly discuss some relevant details about the solution. The eigenfunctions in coordinate space of the Hamiltonian are given explicitly by
\begin{equation}
\langle q| n\rangle= \left(\dfrac{\Omega}{Z} \right)^{1/4} \chi_n\left( q \sqrt{\dfrac{\Omega}{Z}}\right) \exp\left( \dfrac{-i Y q^2}{2Z}\right),
\end{equation}
where we have defined $\Omega=\sqrt{XZ-Y^2}$ and where the functions $\chi_n (x)$ are given by
\begin{equation}
\chi_n(x) = \dfrac{1}{\sqrt{2^n \sqrt{\pi} n!}} e^{-x^2/2} H_n(x)\,,
\end{equation}
with $H_n(x)$ the Hermite polynomials. In this manner, the ground state eigenfunction of the system is given by
\begin{equation}
\langle q| 0\rangle =\left(\frac{\Omega}{\pi\,Z}\right)^{1/4}\exp \left(-\frac{q^{2}}{2Z}\left(\Omega+iY\right)\right)\,.
\end{equation}

If we now perform the following canonical transformation 
\begin{equation}
\label{CanonicalT}
q=\sqrt{Z}Q\,,\hspace{25pt}p=\dfrac{1}{Z}\left(P-YQ \right)\,,
\end{equation}
then the Hamiltonian (\ref{GHO}) takes the standard form of a harmonic oscillator of frequency $\Omega$
\begin{equation}
\label{GHOII}
\widetilde{H}= P^2 + \Omega^2 Q^2\,,
\end{equation}
from where we clearly see that the system's eigenvalues should be 
\begin{equation}
E_n = 2 \,\Omega \left( n+ \frac{1}{2}\right).
\end{equation}
 Moreover, having written the Hamiltonian in this standard form it also becomes clear that position and momentum operators $Q$ and $P$ can be expressed in terms of annihilation and creation operators in the usual manner as
\begin{equation}
\label{CreatAnih}
Q=\sqrt{\frac{1}{2\Omega}}\left(a^{\dagger}+a\right)\,,\hspace{25pt} P=i\,\sqrt{\frac{\Omega}{2}}\left(a^{\dagger}-a\right)\,.
\end{equation}
These operators $a$ and $a^{\dagger}$ act on the eigenstates $\widetilde{|n\rangle}$ of Hamiltonian $\widetilde{H}$, (\ref{GHOII}), in the usual manner. We note the difference between eigenstates $|n\rangle$ of the original Hamiltonian $H$, (\ref{GHO}), and the eigenstates $\widetilde{|n\rangle}$ of the final Hamiltonian $\tilde{H}$, (\ref{GHOII}). Both operators and states are made equivalent by the unitary transformation associated with the canonical change of coordinates (\ref{CanonicalT}).

In the case of the generalized harmonic oscillator the quantum information metric with respect to the ground state can be computed exactly and is given by
\begin{eqnarray}
\label{GHOmet}
g_{ab}  = \dfrac{1}{32\,\Omega^{4}}
\left(
\begin{array}{ccc}
Z^2 & - 2 ZY & 2 Y^2 - XZ  \\
- 2 ZY & 4 XZ & - 2 XY \\
2 Y^2 - XZ & - 2 XY & X^2
\end{array}
\right)\,,
\end{eqnarray}

Moreover, the generalized harmonic oscillator is a system that exhibits a Berry curvature. Its Berry curvature can be exactly computed from the eigenfunctions of the Hamiltonian. (See, e.g. \cite{Chru}.) For the case of the ground state it is given explicitly as a differential form as 
\begin{equation}
\label{F1}
F= \frac{1}{16\,\Omega^{3}}\left(X dY\wedge dZ + Y dZ\wedge dX + Z dX\wedge dY\right)\,,
\end{equation}
where the two-form $F$ presented above is related to the Berry curvature components $F_{ab}$ as given by (\ref{ImZ}) as (see (\ref{2F}) in Appendix.)
\begin{equation}
F = \frac{1}{2}F_{ab}\,dX^{a}\wedge dX^{b}\,.
\end{equation}

\subsection{Application of the Lagrangian Formalism}

In the following subsections we will compute the quantum information metric and the Berry curvature from the two-point expressions (\ref{ReZ}) and (\ref{ImZ}). We will focus on deformations of the ground state $|0\rangle$. To accomplish this we will make use of the annihilation/creation operators to compute the necessary expected values and two-point functions. So, for example, in order to compute the $F_{ZX}$ component of the Berry curvature, we will need to first calculate the following two point function
\begin{equation}
\langle 0 |\mathcal{O}_{Z}(t_{1})\mathcal{O}_{X}(t_{2})|0\rangle=\langle 0 |p^{2}(t_{1})q^{2}(t_{2})|0\rangle\,,
\end{equation}
where we have used expression (\ref{Ox})-(\ref{Oz}) for the deformation operators. Using the transformation (\ref{CanonicalT}), this last expression can also be written as
\begin{equation}
\langle 0 |p^{2}(t_{1})q^{2}(t_{2})|0\rangle=\widetilde{\langle 0 |}\left(\frac{P(t_{1})-Y Q(t_{1})}{\sqrt{Z}}\right)^{2}\left(\sqrt{Z}Q(t_{2})\right)^{2}\widetilde{|0\rangle}\,,
\end{equation}
where we have unitarily transformed both operators and states. This is done so that, following (\ref{CreatAnih}), we can now express the position and momentum operators $Q$ and $P$ in terms of creation and annihilation operators $a$ and $a^{\dagger}$ so that they can act on the eigenstates $\widetilde{|n\rangle}$ in the usual manner.

\subsection{Quantum Information Metric of the Generalized Harmonic Oscillator from the Lagrangian Formalism}

In this section we will focus on the calculation of the quantum information matrix $g_{ab}$ following the expression (\ref{ReZ}). By direct computation in the manner described above, we will first obtain the following expectation values
\begin{eqnarray}
\label{ExpectOx}
\langle 0| \mathcal{O}_{X}|0 \rangle &=&-\langle 0|q^{2}|0\rangle = -\dfrac{Z}{2 \Omega}\,,\\
\label{ExpectOy}
\left<0| \mathcal{O}_Y|0 \right> &=& -\langle 0|\left\{q,p\right\}|0\rangle=\dfrac{Y}{ \Omega}\,,\\
\label{ExpectOz}
\left<0| \mathcal{O}_Z |0\right> &=&-\langle 0|p^{2}|0\rangle= -\dfrac{X}{2 \Omega}\,.
\end{eqnarray}
Having calculated these quantities, we can now concentrate on the computation of the two-point functions needed for each metric element.

\textbullet\; \textbf{Quantum Information Metric Component $g_{XY}$.} We will focus on the computation of the $g_{XY}$ element of the quantum information metric, which according to expression (\ref{ReZ}) is given by
\begin{equation}
\label{gXY}
g_{XY}=\int_{-\infty}^{0}d\tau_{1}\int_{0}^{\infty}d\tau_{2}\;\left(\frac{1}{2}\langle\left\{\mathcal{O}_{X}(\tau_{1})\,,\mathcal{O}_{Y}(\tau_{2})\right\}\rangle-\langle\mathcal{O}_{X}(\tau_{1})\rangle\langle \mathcal{O}_{Y}(\tau_{2})\rangle\right)\,,
\end{equation}
where, according to our identification (\ref{Ox})-(\ref{Oz}), $\mathcal{O}_{X}=-q^{2}$, $\mathcal{O}_{Y}=-\left\{q,p\right\}$. Computing the two-point function $\langle 0 |\mathcal{O}_X(\tau_1) \mathcal{O}_Y(\tau_2)|0 \rangle$ in Wick-rotated time we find
\begin{equation}
\label{OXOY}
\left<0| \mathcal{O}_X(\tau_1) \mathcal{O}_Y(\tau_2)|0 \right> = -\frac{Y Z}{2\Omega^{2}}\left(1+2e^{-4\Omega (\tau_{2}-\tau_{1})}\right)+ i \frac{Z}{\Omega}e^{-4\Omega (\tau_{2}-\tau_{1})}\,,
\end{equation}
while for $\langle 0 |\mathcal{O}_Y(\tau_1) \mathcal{O}_X(\tau_2)|0 \rangle$ we get
\begin{equation}
\label{OYOX}
\left<0| \mathcal{O}_Y(\tau_1) \mathcal{O}_X(\tau_2)|0 \right> = -\frac{Y Z}{2\Omega^{2}}\left(1+2e^{-4\Omega (\tau_{2}-\tau_{1})}\right) -i \frac{Z}{\Omega}e^{-4\Omega (\tau_{2}-\tau_{1})}\,,
\end{equation}
and, using the expressions  (\ref{ExpectOx})-(\ref{ExpectOz}) for the expectation values of the deformation operators, we calculate the product
\begin{equation}
\left<0| \mathcal{O}_X(\tau_1) |0\right> \left<0|  \mathcal{O}_Y(\tau_2) |0\right> = - \dfrac{YZ}{2 \Omega^2}\,,
\end{equation}
Substituting  these results in (\ref{gXY}) we have
\begin{equation}
g_{XY}= \int_{-\infty}^{0}d\tau_1  \int_{0}^{\infty}d\tau_2 \left( -\frac{Y Z}{2\Omega^{2}}\left(1+2e^{-4\Omega (\tau_{2}-\tau_{1})}\right)+ \dfrac{YZ}{2 \Omega^2} \right)\,,
\end{equation}
where we notice that the divergent terms in this integral cancel each other, a feature that will be common in all of our examples. After integration we finally obtain
\begin{equation}
g_{XY}= -\dfrac{Y Z}{16 \Omega^4}\,.
\end{equation}

\textbullet\; \textbf{Quantum Information Metric Component $g_{ZX}$.} In this case, according to expression (\ref{ReZ}) we need to compute 
\begin{equation}
\label{gZX}
g_{ZX}=\int_{-\infty}^{0}d\tau_{1}\int_{0}^{\infty}d\tau_{2}\,\left(\frac{1}{2}\langle\left\{\mathcal{O}_{Z}(\tau_{1})\,,\mathcal{O}_{X}(\tau_{2})\right\}\rangle-\langle\mathcal{O}_{Z}(\tau_{1})\rangle\langle\mathcal{O}_{X}(\tau_{2})\rangle\right)\,,
\end{equation}
with $\mathcal{O}_{Z}=-p^{2}$, $\mathcal{O}_{X}=-q^{2}$. Computing the relevant two-point function in Wick-rotated time we have
\begin{equation}
\label{OZOX}
\left<0| \mathcal{O}_Z(\tau_1) \mathcal{O}_X(\tau_2) |0\right> =  \dfrac{1}{2 \Omega^2}\left(\frac{XZ}{2} + \left(2 Y^2 - XZ\right)e^{-4 \Omega(\tau_2 - \tau_1)}\right)+ i \left(\frac{Y}{\Omega}\right)e^{-4\Omega (\tau_{2}-\tau_{1})}\,,
\end{equation}
while for the element $\langle 0 | \mathcal{O}_{X}(\tau_{1})\mathcal{O}_{Z}(\tau_{2})|0\rangle$
\begin{equation}
\label{OXOZ}
\left<0| \mathcal{O}_X(\tau_1) \mathcal{O}_Z(\tau_2) |0\right> =  \dfrac{1}{2 \Omega^2}\left(\frac{XZ}{2} + \left(2 Y^2 - XZ\right)e^{-4 \Omega(\tau_2 - \tau_1)}\right)- i \left(\frac{Y}{\Omega}\right)e^{-4\Omega (\tau_{2}-\tau_{1})}\,.
\end{equation}
For the product of expectation values we have
\begin{equation}
\left<0| \mathcal{O}_Z(\tau_1)|0 \right> \left<0| \mathcal{O}_X(\tau_2) |0\right> =  \dfrac{XZ}{4 \Omega^2}\,.
\end{equation}
Substituting these results in (\ref{gZX}) and integrating we finally obtain
\begin{equation}
g_{ZX} = \dfrac{2 Y^2 - XZ}{32\, \Omega^4}\,.
\end{equation}

\textbullet\; \textbf{Quantum Information Metric Component $g_{YZ}$.} In this particular case we need to compute
\begin{equation}
\label{gYZ}
g_{YZ}=\int_{-\infty}^{0}d\tau_{1}\int_{0}^{\infty}d\tau_{2}\,\left(\frac{1}{2}\langle\left\{\mathcal{O}_{Y}(\tau_{1})\,,\mathcal{O}_{Z}(\tau_{2})\right\}\rangle-\langle\mathcal{O}_{Y}(\tau_{1})\rangle\langle\mathcal{O}_{Z}(\tau_{2})\rangle\right)\,,
\end{equation}
with $\mathcal{O}_{Y}=-\left\{q,p\right\}$, $\mathcal{O}_{Z}=-p^{2}$. Computing the relevant two-point function in Wick-rotated time we have
\begin{equation}
\label{OYOZ}
\left<0| \mathcal{O}_Y(\tau_1) \mathcal{O}_Z(\tau_2) |0 \right> =    - \dfrac{X Y}{2 \Omega^2}\left( 1+2 e^{- 4 \Omega(\tau_2-\tau_1)}   \right)+i\left(\dfrac{X}{\Omega}\right)e^{- 4 \Omega(\tau_2-\tau_1)}\,,
\end{equation}
while for the element $\left<0| \mathcal{O}_Z(\tau_1) \mathcal{O}_Y(\tau_2) |0\right>$ 
\begin{equation}
\label{OZOY}
\left<0| \mathcal{O}_Z(\tau_1) \mathcal{O}_Y(\tau_2) |0 \right> =    - \dfrac{X Y}{2 \Omega^2}\left( 1+2 e^{- 4 \Omega(\tau_2-\tau_1)}   \right)-i\left(\dfrac{X}{\Omega}\right)e^{- 4 \Omega(\tau_2-\tau_1)}\,,
\end{equation}
For the product of expectation values we obtain
\begin{equation}
\left<0| \mathcal{O}_Y(\tau_1) |0\right> \left<0|  \mathcal{O}_Z(\tau_2)|0 \right> = - \dfrac{X Y}{2 \Omega^2}\,.
\end{equation}
Taking these results, substituting in (\ref{gYZ}) and integrating we have
\begin{equation}
g_{YZ} = -\dfrac{ X Y}{16\, \Omega^4}\,.
\end{equation}

\textbullet\; \textbf{Quantum Information Metric Component $g_{XX}$.} We will now focus on the diagonal terms of the metric. To obtain the $g_{XX}$ element, we must compute 
\begin{equation}
\label{gXX}
g_{XX}=\int_{-\infty}^{0}d\tau_{1}\int_{0}^{\infty}d\tau_{2}\,\left(\langle\mathcal{O}_{X}(\tau_{1})\mathcal{O}_{X}(\tau_{2})\rangle-\langle\mathcal{O}_{X}(\tau_{1})\rangle\langle\mathcal{O}_{X}(\tau_{2})\rangle\right)\,,
\end{equation}
with $\mathcal{O}_{X}=-q^{2}$. If we first compute the relevant two-point function in Wick-rotated time 
\begin{equation}
\left<0| \mathcal{O}_X(\tau_1) \mathcal{O}_X(\tau_2) |0\right> =  \dfrac{Z^2}{4 \Omega^2}\left( 1 +2e^{- 4 \Omega(\tau_2 - \tau_1)} \right)\,,
\end{equation}
and the product of expectation values
\begin{equation}
\left<0| \mathcal{O}_X(\tau_1)|0 \right> \left<0|  \mathcal{O}_X(\tau_2)|0 \right> =  \dfrac{Z^2}{4 \Omega^2}\,,
\end{equation}
then, substituting these results in (\ref{gXX}) we obtain
\begin{equation}
g_{XX} =  \dfrac{Z^2}{32 \Omega^4}\,.
\end{equation}

\textbullet\; \textbf{Quantum Information Metric Component $g_{YY}$.} Similarly, for the $g_{YY}$ element we must compute
\begin{equation}
\label{gYY}
g_{YY}=\int_{-\infty}^{0}d\tau_{1}\int_{0}^{\infty}d\tau_{2}\,\left(\langle\mathcal{O}_{Y}(\tau_{1})\mathcal{O}_{Y}(\tau_{2})\rangle-\langle\mathcal{O}_{Y}(\tau_{1})\rangle\langle\mathcal{O}_{Y}(\tau_{2})\rangle\right)\,,
\end{equation}
with $\mathcal{O}_{Y}=-\left\{ q,p\right\}$. Computing the relevant two-point function in Wick-rotated time we obtain
\begin{equation}
\left<0| \mathcal{O}_Y(\tau_1) \mathcal{O}_Y(\tau_2) |0\right> =\frac{1}{\Omega^{2}}\left(  Y^2 + 2XZe^{- 4 \Omega(\tau_2 - \tau_1)}\right)\,,
\end{equation}
and for the product of expectation values we have
\begin{equation}
\left<0| \mathcal{O}_Y(\tau_2) |0\right> \left<0|  \mathcal{O}_Y(\tau_1) |0\right> =  \dfrac{Y^2}{ \Omega^2}\,,
\end{equation}
so that, substituting in (\ref{gYY}) we obtain
\begin{equation}
g_{YY} =  \dfrac{XZ}{8\,\Omega^4}\,,
\end{equation}

\textbullet\; \textbf{Quantum Information Metric Component $g_{ZZ}$.} Finally, for the $g_{ZZ}$ element we must compute
\begin{equation}
\label{gZZ}
g_{ZZ}=\int_{-\infty}^{0}d\tau_{1}\int_{0}^{\infty}d\tau_{2}\,\left(\langle\mathcal{O}_{Z}(\tau_{1})\mathcal{O}_{Z}(\tau_{2})\rangle-\langle\mathcal{O}_{Z}(\tau_{1})\rangle\langle\mathcal{O}_{Z}(\tau_{2})\rangle\right)\,.
\end{equation}
with $\mathcal{O}_{Z}=-p^{2}$. Calculating the needed two-point function expression in Wick-rotated time we have
\begin{equation}
\left<0| \mathcal{O}_Z(\tau_1) \mathcal{O}_Z(\tau_2) |0\right> =  \frac{X^{2}}{4\Omega^{2}}\left(1+ 2 e^{- 4 \Omega(\tau_1 - \tau_2)} \right)\,,
\end{equation}
and for the product of expected values we obtain
\begin{equation}
\left<0| \mathcal{O}_Z(\tau_2) |0\right> \left< 0| \mathcal{O}_Z(\tau_1)|0 \right> =  \dfrac{X^2}{4 \Omega^2}\,,
\end{equation}
so that, substituting these results in (\ref{gZZ}) we have
\begin{equation}
g_{ZZ} =  \dfrac{X^2}{32\, \Omega^4}\,.
\end{equation}

\bigskip

Having calculated all the quantum information matrix elements above, we can finally write $g_{ab}$ as
\begin{eqnarray}
g_{ab}  = \dfrac{1}{32\,\Omega^{4}}
\left(
\begin{array}{ccc}
Z^2 & - 2 ZY & 2 Y^2 - XZ  \\
- 2 ZY & 4 XZ & - 2 XY \\
2 Y^2 - XZ & - 2 XY & X^2
\end{array}
\right)\,,
\end{eqnarray}
which is exactly the same result one obtains by direct computation (\ref{GHOmet}).

\subsection{Berry Curvature of the Generalized Harmonic Oscillator from the Lagrangian Formalism}

In this section we will describe the computation of the elements of the Berry curvature $F_{ab}$ following the Lagrangian formalism. Generally speaking, the diagonal terms of the tensor are zero because of its antisymmetric character. We can then focus our attention on the remaining three independent components $F_{ZX}$, $F_{XY}$ and $F_{YZ}$. 

\bigskip

\textbullet\; \textbf{Berry Curvature Component $F_{XY}$.} We will compute the component $F_{XY}$, which according to the Lagrangian formalism is given by 
\begin{equation}
\label{FXY}
F_{XY}=\frac{1}{i}\int_{-\infty}^{0}d\tau_{1}\int_{0}^{\infty}d\tau_{2}\,\langle\left[\mathcal{O}_{X}(\tau_{1})\,,\mathcal{O}_{Y}(\tau_{2})\right]\rangle\,,
\end{equation}
with $\mathcal{O}_{X}=-q^{2}$ and $\mathcal{O}_{Y}=-\left\{q,p\right\}$. Using the expressions (\ref{OXOY}) and (\ref{OYOX}) for $\langle 0|\mathcal{O}_{X}(\tau_{1})\mathcal{O}_{Y}(\tau_{2})|0\rangle$ and $\langle 0|\mathcal{O}_{Y}(\tau_{1})\mathcal{O}_{X}(\tau_{2})|0\rangle$ and commutating, we obtain
\begin{equation}
\frac{1}{i}\left(\langle 0|\mathcal{O}_{X}(\tau_{1})\mathcal{O}_{Y}(\tau_{2})|0\rangle-\langle 0|\mathcal{O}_{Y}(\tau_{1})\mathcal{O}_{X}(\tau_{2})|0\rangle\right)=\frac{2Z}{\Omega}e^{-4\Omega(\tau_{2}-\tau_{1})}\,.
\end{equation}
Substituting in this result in (\ref{FXY}) and integrating, we obtain
\begin{equation}
F_{XY}=\frac{Z}{8\Omega^{3}}\,.
\end{equation}

\textbullet\; \textbf{Berry Curvature Component $F_{ZX}$.} In order to obtain the $F_{ZX}$ component of the Berry curvature, according to expression (\ref{ImZ}) we must calculate
\begin{equation}
\label{FZX}
F_{ZX}=\frac{1}{i}\int_{-\infty}^{0}d\tau_{1}\int_{0}^{\infty}d\tau_{2}\,\langle\left[\mathcal{O}_{Z}(\tau_{1})\,,\mathcal{O}_{X}(\tau_{2})\right]\rangle\,,
\end{equation}
where $\mathcal{O}_{X}=-q^{2}$ and $\mathcal{O}_{Z}=-p^{2}$. Using the expressions (\ref{OZOX}) and (\ref{OXOZ}) for $\langle 0|\mathcal{O}_{Z}(\tau_{1})\mathcal{O}_{X}(\tau_{2})|0\rangle$ and $\langle 0|\mathcal{O}_{X}(\tau_{1})\mathcal{O}_{Z}(\tau_{2})|0\rangle$ and commutating, we have
\begin{equation}
\frac{1}{i}\left(\langle 0|\mathcal{O}_{Z}(\tau_{1})\mathcal{O}_{X}(\tau_{2})|0\rangle-\langle 0|\mathcal{O}_{X}(\tau_{1})\mathcal{O}_{Z}(\tau_{2})|0\rangle\right)=\frac{2Y}{\Omega}e^{-4\Omega(\tau_{2}-\tau_{1})}\,.
\end{equation}
Substituting in this result in (\ref{FZX}) and integrating, we obtain
\begin{equation}
F_{ZX}=\frac{Y}{8\Omega^{3}}\,.
\end{equation}

\textbullet\; \textbf{Berry Curvature Component $F_{YZ}$.} To obtain the $F_{YZ}$ component we must compute
\begin{equation}
\label{FYZ}
F_{YZ}=\frac{1}{i}\int_{-\infty}^{0}d\tau_{1}\int_{0}^{\infty}d\tau_{2}\,\langle\left[\mathcal{O}_{Y}(\tau_{1})\,,\mathcal{O}_{Z}(\tau_{2})\right]\rangle\,,
\end{equation}
with $\mathcal{O}_{Y}=-\left\{q,p\right\}$ and $\mathcal{O}_{Z}=-p^{2}$. Using the expressions (\ref{OYOZ}) and (\ref{OZOY}) for $\langle 0|\mathcal{O}_{Y}(\tau_{1})\mathcal{O}_{Z}(\tau_{2})|0\rangle$ and $\langle 0|\mathcal{O}_{Z}(\tau_{1})\mathcal{O}_{Y}(\tau_{2})|0\rangle$ and commutating, we obtain
\begin{equation}
\frac{1}{i}\left(\langle 0|\mathcal{O}_{Y}(\tau_{1})\mathcal{O}_{Z}(\tau_{2})|0\rangle-\langle 0|\mathcal{O}_{Z}(\tau_{1})\mathcal{O}_{Y}(\tau_{2})|0\rangle\right)=\frac{2X}{\Omega}e^{-4\Omega(\tau_{2}-\tau_{1})}\,.
\end{equation}
Substituting in this result in (\ref{FYZ}) and integrating, we obtain
\begin{equation}
F_{YZ}=\frac{X}{8\Omega^{3}}\,.
\end{equation}

Therefore, the writing the Berry curvature obtained in this manner as a two-form $F=\frac{1}{2}F_{ab}dX^{a}\wedge dX^{b}$ we obtain the following result
\begin{equation}
F = \frac{1}{16 \,\Omega^{3}}\left(X\,dY\wedge dZ + Y\,dZ\wedge dX+Z\,dX\wedge dY\right)
\end{equation}
which is exactly the expression (\ref{F1}) obtained through direct computation. In this manner we can clearly see that the Lagrangian method yields the same results to the quantum information metric and the Berry curvature.


\section{Quantum XY Model in a Transverse Magnetic Field}

\subsection{Computation of the Quantum Information Metric and Berry Curvature in Quantum Spin Chains}

In the next examples we will compute the quantum information metric and the Berry curvature in two different quantum spin chain systems: the quantum XY model in a transverse magnetic field and the generalized Ising model. For general information about quantum spin chains and their solutions, the reader is refered to \cite{Lieb, Katsura, Parkinson}. In the following subsection we will summarily describe the standard procedure when dealing with quantum spin chain systems in the particular example of the XY model.

\subsection{Quantum XY Model in a Transverse Magnetic Field: General Setting}

The main purpose of the present example is to compute the quantum information metric for the quantum XY model. Roughly speaking, the XY model can be seen as a generalization of the Ising model in which an anisotropy is introduced with respect to the $X$ and $Y$ directions by means of a real parameter $\gamma$. The Hamiltonian of the system is given by
\begin{equation} \label{eq:hamxymodsd}
H = - \sum_{l=-M}^M \left[ \left(\dfrac{1+\gamma}{2}\right)\sigma^x_l \sigma^x_{l+1} +\left(\dfrac{1-\gamma}{2}\right)\sigma^y_l \sigma^y_{l+1}+ h\, \sigma^z_l\right]\,,
\end{equation}
where the total odd number of spins is $N=2M+1$ and $h$ is the transverse magnetic field. We note that in the $\gamma=1$ case the systems reduces to a one-dimensional transverse-field Ising model. 

In order to diagonalize Hamiltonian (\ref{eq:hamxymodsd}) we will follow the standard procedure used originally to study the quantum Ising model described in \cite{Lieb, Katsura}. See also \cite{Parkinson, Sachdev} for a more detailed review. Following this procedure, we begin by defining operators $\sigma^{+}$, $\sigma^{-}$ as
\begin{equation}
\sigma^+= \dfrac{1}{2} \left( \sigma^x + i \sigma^y\right), \ \ \ \ \sigma^-=\dfrac{1}{2} \left( \sigma^x - i \sigma^y\right)\,,
\end{equation}
so that the Hamiltonian (\ref{eq:hamxymodsd}) takes the form
\begin{eqnarray}
H=-\sum^M_{l=-M} \left[\dfrac{\gamma+1}{2}\left(\sigma^+_l \sigma^+_{l+1} + \sigma^+_l \sigma^-_{l+1} + \sigma^-_l \sigma^+_{l+1} + \sigma^-_l \sigma^-_{l+1} \right) \right. \nonumber \\
 + \left. \dfrac{\gamma-1}{2} \left(\sigma^+_l \sigma^+_{l+1} - \sigma^+_l \sigma^-_{l+1} - \sigma^-_l \sigma^+_{l+1}+ \sigma^-_l \sigma^-_{l+1} \right) + h\sigma^z_l \right]\,.
\end{eqnarray}

We will now perform Jordan-Wigner transformation, which consists in relating the spin operators $\sigma^{+}_{l}$, $\sigma^{-}_{l}$ and $\sigma^{z}_{l}$ to a set of fermionic operators $a_{l}$ and $a^{\dagger}_{l}$ by proposing
\begin{eqnarray}
\sigma^+_l &=& \left( \prod_{j=1}^{l-1} \sigma_j^z \right) a_l\,,\\
\sigma^-_l &=& \left( \prod_{j=1}^{l-1} \sigma_j^z \right) a^\dagger_l\,,\\
\sigma^z_l &=& 1-2 a_l^\dagger a_l\,,
\end{eqnarray}
where the operators $a_l^\dagger$ and $a_l$ satisfy the anticommutation relations
\begin{equation}
\left\lbrace a^\dagger_l, a_m \right\rbrace = \delta_{l,m}\,,
\end{equation}
\begin{equation}
\left\lbrace a^\dagger_l, a_m^\dagger \right\rbrace = \left\lbrace a_l, a_m \right\rbrace=0\,.
\end{equation}
Having acted in this manner, the Hamiltonian written in terms of the fermionic operators $a_{l}$, $a_{l}^{\dagger}$ is given by
\begin{eqnarray}
H&=&-\sum_{l=-M}^M\left[ \dfrac{\gamma+1}{2}\left(a_{l+1}a_l +a^\dagger_{l+1} a_l + a^\dagger_l a_{l+1}+ a^\dagger_l a^\dagger_{l+1} \right) + \right. \nonumber \\
&& \left.  \dfrac{\gamma-1}{2}\left(a_{l+1}a_l -a^\dagger_{l+1} a_l - a^\dagger_l a_{l+1}+ a^\dagger_l a^\dagger_{l+1} \right) +h \left(1-2 a^\dagger_l a_l \right)\right]\,.
\end{eqnarray}
We now apply a Fourier transform as 
\begin{equation}
\label{FourierOp}
a_l = \dfrac{1}{\sqrt{N}} \sum_k e^{- i k l} d_k\,, \ \ \ \ \ a^\dagger_l = \dfrac{1}{\sqrt{N}} \sum_k e^{i k l} d^\dagger_k\,,
\end{equation}
where $k=   2 \pi/N, 4 \pi/N, ...,  2 \pi $. Making use of the fact that the Delta function can be expressed as
\begin{equation}
\delta_{k,k'} = \dfrac{1}{N} \sum_l e^{i l(k-k')}\,,
\end{equation}
then the Hamiltonian can be reduced to the form
\begin{equation}
\label{HXYd}
H=-\sum_k\left[  2\left(  \cos k-  h \right)d_k^\dagger d_k - i \gamma \sin k  \left( d_k d_{-k} + d^\dagger_k d^\dagger_{-k} \right) \right] - h N\,.
\end{equation}

Finally, we now apply a Bogoliubov transformation given by
\begin{eqnarray} \label{eq:daniq}
d_k &=& \cos\frac{\theta_k}{2}\,b_k + i \sin\frac{\theta_k}{2}\,b_{-k}^\dagger\,,\\
 \label{eq:ddagcr}
d_k^\dagger &=& \cos  \dfrac{\theta_k}{2}\, b^\dagger_k - i \sin  \dfrac{\theta_k}{2}\, b_{-k}\,.
\end{eqnarray}

We still need to determine the functional dependence of $\theta_{k}$ on the parameters $h$ and $k$. Substituting the transformation (\ref{eq:daniq}) and (\ref{eq:ddagcr}) in the Hamiltonian (\ref{HXYd}) we obtain
\begin{eqnarray}
H &=& -\sum_{k}2\left(\gamma \sin k \sin \theta_{k} b^{\dagger}_{k}b_{k} - (\cos k -h) \cos \theta_{k} b^{\dagger}_{k}b_{k}\right) \nonumber \\
&& -i \sum_{k}\left( (\cos k -h) \sin \theta_{k}-  \gamma \sin k \cos\theta_{k} \right)\left(b_{k}b_{-k}+ b^{\dagger}_{k}b^{\dagger}_{-k}\right)+\textit{const.}\,,
\end{eqnarray}
and imposing that the cross-terms be zero we arrive at the following equation
\begin{equation}
 (\cos k-h )\sin \theta_{k}- \gamma\sin k\cos \theta_k =0\,.
\end{equation}
from where we find that
\begin{eqnarray}
\cos \theta_k &=&\dfrac{\cos k -h}{\sqrt{\left(\cos k -h \right)^2+ \gamma^2 \sin^2 k}}\,,\\
\label{SinTheta}
\sin\theta_k &=& -\dfrac{\gamma \sin k}{\sqrt{\left(\cos k -h \right)^2+ \gamma^2 \sin^2 k}}\,,
\end{eqnarray}
and we find that the Hamiltonian can be written as a quasi-free fermionic system
\begin{equation}
\label{XYDiag}
H=\sum_k \Lambda_k \left( b_k^\dagger b_k-1 \right)\,,
\end{equation}
where we have defined the quasi-particle dispersion relation $\Lambda_{k}$ as 
\begin{equation}
\label{Lk}
\Lambda_k = \sqrt{\left(\cos k-h \right)^2+ \gamma^2 \sin^2 k}\,.
\end{equation}

\bigskip

Having put the Hamiltonian in the diagonal form (\ref{XYDiag}), we can now find the explicit form of the ground-state $|\Omega\rangle$. This state is defined such that for any annihilation operator $b_{k}$ we have
\begin{equation}
\label{GSCond}
b_{k}|\Omega\rangle = 0\,.
\end{equation}
Clearly then, by acting the Hamiltonian (\ref{XYDiag}) on $|\Omega\rangle$ we obtain the ground state energy $E_{0}=-\sum_{k}\Lambda_{k}$. Similarly, we will define an excited state $|k,-k\rangle$ as
\begin{equation}
|k,-k\rangle = b^{\dagger}_{k}b^{\dagger}_{-k}|\Omega\rangle\,,
\end{equation}
with energy $E_{k,-k}$ given by
\begin{equation}
E_{k,-k}=2\Lambda_{k}-	\sum_{k'} \Lambda_{k'}\,.
\end{equation}

In order to find the explicit form of the ground state $|\Omega\rangle$ we first divide it in its different modes $|\Omega\rangle =\prod_{k}|\Omega\rangle_{k}$ and make the following ansatz for each one of them
\begin{equation}
\left\vert \Omega\right>_k = a \left\vert 0\right>_k \left\vert 0\right>_{-k} +b \left\vert 1\right>_k \left\vert 0\right>_{-k}+c \left\vert 0\right>_k \left\vert 1\right>_{-k}+d \left\vert 1\right>_k \left\vert 1\right>_{-k}\,.
\end{equation}
If we now impose condition (\ref{GSCond}), we arrive at the following results for coefficients $a$, $b$, $c$ and $d$
 \begin{equation}
 a=\cos\frac{\theta_k}{2} \,,\hspace{15pt}\ \ d=- i \sin\dfrac{\theta_k}{2} \,, \hspace{15pt}\ \ c=b=0,
 \end{equation}
so the complete ground state of the system is given by
 \begin{equation}
 \label{GroundXY}
 \left\vert \Omega \right> = \prod_k\left( \cos \frac{\theta_k}{2}  \left\vert 0 \right>_k  \left\vert 0 \right>_{-k} - i \sin \frac{\theta_k}{2} \left\vert 1 \right>_k  \left\vert 1 \right>_{-k} \right)\,.
 \end{equation}
 
\bigskip

The quantum information metric for this system is known \cite{Zanardi_2006, Zanardi} and its elements are given explicitly as 
\begin{eqnarray}
\label{ghh}
g_{hh}&=& \frac{1}{4}\sum_{k}\frac{\gamma^{2}\sin^{2}k}{\left[\left(\cos k - h\right)^{2}+\gamma^{2}\sin^{2}k\right]^{2}}\,,\\
\label{ggammagamma}
g_{\gamma\gamma}&=& \frac{1}{4}\sum_{k}\frac{\sin^{2} k \left(\cos k -h\right)^{2}}{\left[\left(\cos k - h\right)^{2}+\gamma^{2}\sin^{2}k\right]^{2}}\,,\\
\label{ghgamma}
g_{h\gamma}&=&-\frac{\gamma}{4}\sum_{k}\frac{ \sin^{2}k\left(\cos k - h\right)}{\left[\left(\cos k - h\right)^{2}+\gamma^{2}\sin^{2}k\right]^{2}}\,.
\end{eqnarray}

 \subsection{Quantum Information Metric of the Quantum XY Model from the Lagrangian Formalism}
 
In this section we will compute the quantum information metric with respect to the ground state using the Lagrangian formalism and compare with expressions (\ref{ghh})-(\ref{ghgamma}) found in the literature.

The first task is to find the explicit form of the deformation operators related to changes in the parameters of the system. By inspection of the XY model Hamiltonian (\ref{eq:hamxymodsd}) we can directly identify the deformation operator $\mathcal{O}_{h}$ related to changes in the value of the magnetic field $h$. This operator is given explicitly by
 \begin{equation}
 \mathcal{O}_h = \sum_l \sigma_l^z\,.
 \end{equation}
If we now apply the transformations used in the previous section to diagonalize the Hamiltonian to its form (\ref{XYDiag}), the operator $\mathcal{O}_{h}$ can be written as
 \begin{equation} \label{eq:operah}
 \mathcal{O}_h = N - 2 \sum_k \left[ \cos^2 \frac{\theta_k}{2}\;b^\dagger_k b_k + \sin^2  \frac{\theta_k}{2}\;b_k b^\dagger_k + \frac{i}{2}\sin \theta_k\left( b^\dagger_k b^\dagger_{-k} + b_k b_{-k}\right)\right]\,.
 \end{equation}
 
In a similar fashion, starting from the XY model Hamiltonian (\ref{eq:hamxymodsd}) we can also directly identify the operator $\mathcal{O}_{\gamma}$ associated with a change on the anisotropy parameter $\gamma$. It is given explicitly by
\begin{equation}
\mathcal{O}_\gamma = \frac{1}{2} \sum_l \left( \sigma^x_l \sigma^x_{l+1}- \sigma^y_l \sigma^y_{l+1}\right)\,.
\end{equation}
If we now again apply to it the transformations described in the previous section for the diagonalization of Hamiltonian (\ref{eq:hamxymodsd}), then $\mathcal{O}_{\gamma}$ is written as
\begin{equation} \label{eq:opgamdiag}
\mathcal{O}_\gamma = -i \sum_k \sin k \left[\cos\theta_k \left( b_k b_{-k}+b_k^\dagger b^\dagger_{-k}\right) +i \sin\theta_k\left( b^\dagger_k b_k - b_k b^\dagger_k\right) \right]\,.
\end{equation}

\textbullet\; \textbf{Quantum Information Metric Component $g_{hh}$.} Having determined the explicit form of operator $\mathcal{O}_{h}$, we will now focus our attention on the computation of the quantum information matrix element $g_{hh}$ following the Lagrangian formalism. According to this formalism, the $g_{hh}$ element with respect to the ground state $|\Omega\rangle$ is given by
\begin{equation}
\label{ghhL}
g_{hh} = \int_{-\infty}^{0}d\tau_{1}\int_{0}^{\infty}d\tau_{2}\;\left(\langle \Omega | \mathcal{O}_{h}(\tau_{1})\mathcal{O}_{h}(\tau_{2})| \Omega\rangle-\langle \Omega | \mathcal{O}_{h}(\tau_{1}) |\Omega \rangle \langle \Omega | \mathcal{O}_{h}(\tau_{2}) |\Omega \rangle   \right)\,,
\end{equation}
so that, in order to calculate this expression we will need to compute the relevant expectation values and two-point functions. We proceed by first computing the expected value $\langle \Omega | \mathcal{O}_{h} | \Omega \rangle$. We do this by taking advantage that the operator $\mathcal{O}_{h}$ as given by expression (\ref{eq:operah}) is written in terms of operators $b_{k}$ and $b_{k}^{\dagger}$ that act directly on the ground state $| \Omega \rangle$. The final result is given by
 \begin{equation}
 \label{ExpOh}
 \left<\Omega| \mathcal{O}_h(\tau)|\Omega\right> = N - 2 \sum_k \sin^2  \frac{\theta_k}{2}\,.
 \end{equation}
 Acting in a similar manner, for the required two-point function $\left<\Omega| \mathcal{O}_h(t_1) \mathcal{O}_h(t_2)|\Omega\right>$ we obtain
\begin{eqnarray}
\left<\Omega| \mathcal{O}_h(\tau_1) \mathcal{O}_h(\tau_2)|\Omega\right> &=& N^2 -4 N \sum_k \sin^2 \frac{\theta_k}{2}  +\sum_k  e^{-2\Lambda_{k}(\tau_2-\tau_1)} \sin^2 \theta_k \nonumber \\
&& + 4 \sum_{k,k'} \sin^2 \frac{\theta_k}{2} \sin^2 \frac{\theta_{k'}}{2} \,.
\end{eqnarray}
Substituting these results in expression (\ref{ghhL}) and integrating we obtain
\begin{equation}
g_{hh}= \sum_k \dfrac{\sin^2 \theta_k}{4\Lambda_{k}^2}\,,
\end{equation}
or, substituting the expression for $\sin\theta_k$ from (\ref{SinTheta}) and the explicit form (\ref{Lk}) for $\Lambda_{k}$ we finally have
\begin{equation}
g_{hh}= \frac{1}{4} \sum_k \dfrac{\gamma^2 \sin^2 k}{\left[\left(\cos k-h \right)^2+ \gamma^2 \sin^2 k\right]^2}\,,
\end{equation}
which coincides exactly with the expression (\ref{ghh}) found in the literature.
\bigskip

\textbullet\; \textbf{Quantum Information Metric Component $g_{\gamma \gamma}$.} We will now compute the component $g_{\gamma\gamma}$ of the quantum information metric following the Lagrangian formalism. Accordingly, we must compute
\begin{equation}
\label{ggamgam}
g_{\gamma\gamma} = \int_{-\infty}^{0}d\tau_{1}\int_{0}^{\infty}d\tau_{2}\;\left(\langle \Omega | \mathcal{O}_{\gamma}(\tau_{1})\mathcal{O}_{\gamma}(\tau_{2})| \Omega\rangle-\langle \Omega | \mathcal{O}_{\gamma}(\tau_{1}) |\Omega \rangle \langle \Omega | \mathcal{O}_{\gamma}(\tau_{2}) |\Omega \rangle   \right)\,.
\end{equation}
Once we have written the operator $\mathcal{O}_\gamma$ in the form (\ref{eq:opgamdiag}), we can directly calculate the required expectation values and two-point functions  with respect to the ground state. Proceeding in this manner, for the expected value $\langle \Omega | \mathcal{O}_{\gamma}(\tau)|\Omega\rangle$ we find
\begin{equation}
\label{ExpOg}
\left<\Omega| \mathcal{O}_\gamma (\tau) |\Omega\right> = - \sum_k \sin k \sin \theta_k\,,
\end{equation}
and similarly, for the two-point function we obtain
\begin{eqnarray}
\left<\Omega| \mathcal{O}_\gamma(\tau_1)  \mathcal{O}_\gamma(\tau_2)|\Omega \right> &=&  \sum_k e^{-2\Lambda_{k}(\tau_2-\tau_1)} \sin^2 k\sin^2 \theta_k \nonumber \\
&& + \sum_{k,k'} \sin k \sin k' \sin\theta_k \sin\theta_{k'}\,.
\end{eqnarray}
Substituting these results in the expression (\ref{ggamgam}), integrating and using the expressions (\ref{SinTheta}) and (\ref{Lk}) for $\sin \theta_{k}$ and $\Lambda_{k}$ we obtain the following result 
\begin{equation}
g_{\gamma \gamma} = \frac{1}{4} \sum_k \frac{\sin^2 k\;(\cos k-h)^2}{\left[\left(\cos k-h \right)^2+ \gamma^2 \sin^2 k\right]^2}\,,
\end{equation}
which is exactly the same expression (\ref{ggammagamma}) for $g_{\gamma\gamma}$ found in the literature.

\bigskip

\textbullet\; \textbf{Quantum Information Metric Component $g_{h \gamma}$.}  According to the Lagrangian formalism, this component of the metric is given by
\begin{equation}
\label{ghgam}
g_{h\gamma} = \int_{-\infty}^{0}d\tau_{1}\int_{0}^{\infty}d\tau_{2}\;\left(\frac{1}{2}\langle \Omega | \left\{\mathcal{O}_{h}(\tau_{1})\,,\mathcal{O}_{\gamma}(\tau_{2})\right\}| \Omega\rangle-\langle \Omega | \mathcal{O}_{h}(\tau_{1}) |\Omega \rangle \langle \Omega | \mathcal{O}_{\gamma}(\tau_{2}) |\Omega \rangle   \right)\,.
\end{equation}
Finally, by using the operators (\ref{eq:operah}) and (\ref{eq:opgamdiag}), we can compute the function
\begin{eqnarray}
\left<\Omega| \mathcal{O}_h(\tau_1)  \mathcal{O}_\gamma(\tau_2)|\Omega \right> &=& \sum_k e^{-2\Lambda_{k}(\tau_2 - \tau_1)} \sin k \cos \theta_k \sin \theta_k  -N \sum_k  \sin k \sin \theta_k \nonumber \\
&& + 2 \sum_{k,k'} \sin k \sin \theta_k  \sin^2 \dfrac{\theta_{k'}}{2}\,,
\end{eqnarray}
where we note that there is no imaginary part, and therefore there is no need to anticommutate. Also, using the expressions (\ref{ExpOh}) and (\ref{ExpOg}) for the expected values of $\mathcal{O}_{h}$ and $\mathcal{O}_{\gamma}$ we find that their product is
\begin{equation}
\langle \Omega | \mathcal{O}_{h}(\tau_{1}) |\Omega \rangle \langle \Omega | \mathcal{O}_{\gamma}(\tau_{2}) |\Omega \rangle   =2 \sum_{k,k'} \sin k \sin \theta_k  \sin^2 \dfrac{\theta_{k'}}{2} -N \sum_k  \sin k \sin \theta_k \,.
\end{equation}
Then, combining these results we can obtain the remaining quantum information metric element, which is given explicitly as
\begin{equation}
g_{h \gamma} =- \frac{\gamma}{4}\sum_k \dfrac{ \sin^2k\;(\cos k -h)}{\left[\left(\cos k-h \right)^2+ \gamma^2 \sin^2 k\right]^2}\,,
\end{equation}
which is exactly the same as the expression (\ref{ghgamma}) found in the literature.

We notice that in the previous calculations none of the obtained expected values or two-point functions have an imaginary value. Therefore, there is no possibility for a Berry curvature $F_{ab}$. However, there is a generalization of this system which indeed presents a Berry curvature. This system will be considered in the following section.

\section{Ising Model in a Transverse Field Adiabatically Rotated }

\subsection{General Setting }

In this section we will consider a generalization of the Ising model that is known to present both a quantum information metric and Berry and whose values can be calculated explicitly \cite{PachosI, PachosII}. For additional research on this model, the reader is referred to \cite{Zhu, Hamma, B_Basu, Kolodrubetz, GritsevII}.  We start by considering a one-dimensional Ising model
\begin{equation}
\label{IsingI}
H=- \sum_{l = - M}^{M} \left[ \sigma^x_l \sigma^x_{l+1}+ h\, \sigma^{z}_{l}\right]\,.
\end{equation}
Clearly, this system is the $\gamma=1$ case of the XY model studied in the previous section. Following \cite{PachosI, PachosII} we will apply a rotation of $\phi$ around the $z$-coordinate for each spin by means of the unitary operator
\begin{equation}
g(\phi)= \prod_{l=-M}^Me^{i\phi \frac{\sigma^z}{2}}\,.
\end{equation}
Acting this operator on Hamiltonian (\ref{IsingI}) we obtain
\begin{eqnarray}
H(\phi)&=&g(\phi) H\,g^\dagger(\phi) \nonumber \\
& =& -\sum_l \left[\left( \cos\phi\;\sigma^x_l - \sin\phi\;\sigma^y_l \right)\left(\cos\phi\;\sigma^x_{l+1}- \sin\phi\;\sigma^y_{l+1} \right) +h \sigma^z_{l} \right]\,.
\end{eqnarray} 
If we now diagonalize this rotated Hamiltonian following the same procedure described in the previous section for the XY model, we can rewrite it in the form\footnote{In the diagonal Hamiltonian (\ref{HamRot}) we have set the ground-state energy $E_{0}=0$ to match the standard notation in the references.}
\begin{equation}
\label{HamRot}
H= \sum_{k} \Lambda_k\; b_k^\dagger b_k\,,
\end{equation}
where the quasi-particle operators $b_{k}$ and $b^{\dagger}_{k}$ are now given by
\begin{equation}
b_k = d_k \cos\frac{\theta_k}{2}- i e^{2i \phi} d^\dagger_{-k} \sin \frac{\theta_k}{2}\,,
\end{equation}
and where the Fourier-mode operators $d_{k}$ and $d^{\dagger}_{k}$ are defined in the same manner as in (\ref{FourierOp}) and where $\theta_{k}$ is defined as
\begin{equation}
\theta_k = \cos^{-1}\left[\frac{\cos k - h}{\sqrt{1-2 h \cos k + h^2}}\right]\,.
\end{equation}
The dispersion relation $\Lambda_{k}$ is now given by
\begin{equation}
\Lambda_k = \sqrt{1-2 h \cos k+h^2}\,,
\end{equation}
and the ground-state of the system is given explicitly by 
\begin{equation}
\label{IsingGround}
\left\vert \Omega \right> = \prod_k \left(\cos\frac{\theta_k}{2} \left\vert 0 \right>_k \left\vert 0 \right>_{-k}-i e^{2i\phi} \sin\frac{\theta_k}{2}  \left\vert 1 \right>_k\left\vert 1 \right>_{-k}\right)\,,
\end{equation}

The quantum information metric elements have been computed in \cite{Kolodrubetz}. They are given as
\begin{eqnarray}
\label{gLL}
g_{hh}&=& \frac{1}{4}\sum_{k}\frac{\sin^{2}k}{\left[1-2 h \cos k+h^2\right]^{2}}\,,\\
\label{gPhPh}
g_{\phi \phi} &=&\frac{1}{4}\sum_{k}\frac{\sin^{2}k}{\left[1-2 h \cos k+h^2\right]} \,,\\
\label{gPhL}
g_{\phi h} &=& 0\,.
\end{eqnarray}

Likewise, the only independent non-zero element of the Berry curvature $F_{\phi h}$ was also computed in \cite{Kolodrubetz} and is 
\begin{equation}
\label{FPhL}
F_{\phi h}=\frac{1}{2}\sum_{k}\frac{\sin^{2}k}{\left[1-2 h \cos k+h^2\right]^{3/2}}\,.
\end{equation}

\subsection{Quantum Information Metric and Berry Curvature from the Lagrangian Formalism}

The free parameters of the system are given by $\phi$ and $h$, and the deformation operators associated with  changes in these parameters are
\begin{eqnarray}
\label{OL}
\mathcal{O}_h &=& N - \sum_k \left[ i \sin\theta_k \left( e^{2 i \phi}b_k^\dagger b_{-k}^\dagger - e^{-2i \phi}b_{-k}b_{k}\right) + 2\sin^2 \frac{\theta_k}{2} b_{-k} b^\dagger_{-k}+2\cos^{2}\frac{\theta_{k}}{2}b^{\dagger}_{k}b_{k}\right]\,,\nonumber\\
\\
\label{OPh}
\mathcal{O}_\phi &=& - \sum_{k} \Lambda_k \sin\theta_k \left(e^{2i \phi} b^\dagger_k b^\dagger_{-k} +e^{-2i\phi} b_{-k} b_k \right)\,.
\end{eqnarray}

\textbullet\; \textbf{Quantum Information Metric Component $g_{h h}$.} Having computed the explicit form of the deformation operator $\mathcal{O}_{h}$, we will now compute the quantum information metric element $g_{hh}$, which in the Lagrangian formalism is given by
\begin{equation}
\label{gLLLag}
g_{hh} = \int_{-\infty}^{0}d\tau_{1}\int_{0}^{\infty}d\tau_{2}\;\left(\langle \Omega | \mathcal{O}_{h}(\tau_{1})\mathcal{O}_{h}(\tau_{2})| \Omega\rangle-\langle \Omega | \mathcal{O}_{h}(\tau_{1}) |\Omega \rangle \langle \Omega | \mathcal{O}_{h}(\tau_{2}) |\Omega \rangle   \right)\,.
\end{equation}
Using the expression (\ref{OL}) for $\mathcal{O}_{h}$ we can directly compute the two-point function $\langle \Omega | \mathcal{O}_{h}(\tau_{1})\mathcal{O}_{h}(\tau_{2})| \Omega\rangle$. The result is
\begin{eqnarray}
\left< \Omega|\mathcal{O}_h(\tau_1) \mathcal{O}_h(\tau_2)|\Omega\right> &=& N^{2} -4 N\sum_k \sin^2  \frac{\theta_k}{2} +4 \sum_{k,k'}\sin^2 \frac{\theta_k}{2} \sin^2 \frac{\theta_{k'}}{2}\nonumber \\ 
&&+\, \sum_k \sin^2 \theta_k\, e^{-2\Lambda_{k}(\tau_2-\tau_1)}\,,
\end{eqnarray}
where, for the expectation value $ \langle \Omega | \mathcal{O}_{h}(\tau) |\Omega \rangle $ we obtain
\begin{equation}
\left<\Omega| \mathcal{O}_h (\tau)|\Omega\right>=N- 2\sum_k \sin^2 \frac{\theta_k}{2}\,.
\end{equation}
Substituting these results in (\ref{gLLLag}) and integrating, we have the following result for $g_{hh}$
\begin{equation}
g_{h h} = \frac{1}{4}\sum_k \dfrac{\sin^2 k}{\left[1-2 h \cos k+h^2\right]^{2}}\,,
\end{equation}
which is the same expression as (\ref{gLL}) found in the literature \cite{Kolodrubetz}.

\textbullet\; \textbf{Quantum Information Metric Component $g_{\phi \phi}$.} According to the Lagrangian formalism, the $g_{\phi \phi}$ of the quantum information metric is given by
\begin{equation}
\label{gffLag}
g_{\phi\phi} = \int_{-\infty}^{0}d\tau_{1}\int_{0}^{\infty}d\tau_{2}\;\left(\langle \Omega | \mathcal{O}_{\phi}(\tau_{1})\mathcal{O}_{\phi}(\tau_{2})| \Omega\rangle-\langle \Omega | \mathcal{O}_{\phi}(\tau_{1}) |\Omega \rangle \langle \Omega | \mathcal{O}_{\phi}(\tau_{2}) |\Omega \rangle   \right)\,.
\end{equation}
Using the expression (\ref{OPh}) for the operator $\mathcal{O}_{\phi}$ we compute directly the two-point function $\langle \Omega | \mathcal{O}_{\phi}(\tau_{1})\mathcal{O}_{\phi}(\tau_{2})| \Omega\rangle$. The result is
\begin{equation}
\left<\Omega| \mathcal{O}_\phi(\tau_1) \mathcal{O}_\phi(\tau_2)|\Omega\right> = \sum_{k=-M}^M e^{-2\Lambda_{k}(\tau_2-\tau_1)} \Lambda_{k}^2 \sin^2\theta_k\,,
\end{equation}
whereas for the expectation value $\langle \Omega | \mathcal{O}_{\phi}(\tau) |\Omega \rangle$ we obtain
\begin{equation}
\left< \Omega|\mathcal{O}_\phi (\tau)|\Omega\right> =0\,.
\end{equation}
Substituting these results in (\ref{gffLag}), we obtain 
\begin{equation}
g_{\phi \phi}= \frac{1}{4}\sum_{k}\frac{\sin^{2}k}{\left[1-2 h \cos k+h^2\right]}\,,
\end{equation}
which is the exact same expression (\ref{gPhPh}) as found in \cite{Kolodrubetz}.

\textbullet\; \textbf{Quantum Information Metric Component $g_{\phi h}$.} We now focus on the component $g_{\phi h}$, which according to the Lagrangian formalism is given by
\begin{equation}
\label{gfLLag}
g_{\phi h} = \int_{-\infty}^{0}d\tau_{1}\int_{0}^{\infty}d\tau_{2}\;\left(\frac{1}{2}\langle \Omega |\left\{ \mathcal{O}_{\phi}(\tau_{1})\,,\mathcal{O}_{h}(\tau_{2})\right\}| \Omega\rangle-\langle \Omega | \mathcal{O}_{\phi}(\tau_{1}) |\Omega \rangle \langle \Omega | \mathcal{O}_{h}(\tau_{2}) |\Omega \rangle   \right)\,.
\end{equation}
Using the expressions (\ref{OL}) and (\ref{OPh}) for the operators $\mathcal{O}_{h}$ and $\mathcal{O}_{\phi}$, we find for the two-point function $\langle \Omega |\mathcal{O}_{\phi}(\tau_{1})\mathcal{O}_{h}(\tau_{2})| \Omega\rangle$ the following result
\begin{equation}
\label{Ofh}
\langle \Omega |\mathcal{O}_{\phi}(\tau_{1})\mathcal{O}_{h}(\tau_{2})| \Omega\rangle = i 	\sum_{k} \frac{\sin^{2}k}{\sqrt{1-2 h \cos k+h^2}}e^{-2\Lambda_{k}(\tau_{2}-\tau_{1})}\,,
\end{equation}
while for the element $\langle \Omega |\mathcal{O}_{h}(\tau_{1})\mathcal{O}_{\phi}(\tau_{2})| \Omega\rangle$ we find
\begin{equation}
\label{Ohf}
\langle \Omega |\mathcal{O}_{h}(\tau_{1})\mathcal{O}_{\phi}(\tau_{2})| \Omega\rangle =-i 	\sum_{k} \frac{\sin^{2}k}{\sqrt{1-2 h \cos k+h^2}}e^{-2\Lambda_{k}(\tau_{2}-\tau_{1})}\,.
\end{equation}
and we already have found that $\langle \Omega | \mathcal{O}_{\phi}(\tau) |\Omega \rangle=0$. Substituting these results in (\ref{gfLLag}), we then find that
\begin{equation}
g_{\phi h}=0\,,
\end{equation}
just as it was found in (\ref{gPhL}).

\textbullet\; \textbf{Berry Curvature Component $F_{\phi h}$.} Finally, we will now compute the component $F_{\phi h}$ of the Berry curvature. According to the Lagrangian formalism it is given by
\begin{equation}
\label{FfLLag}
F_{\phi h} =\frac{1}{i} \int_{-\infty}^{0}d\tau_{1}\int_{0}^{\infty}d\tau_{2}\;\langle \Omega |\left[ \mathcal{O}_{\phi}(\tau_{1})\,,\mathcal{O}_{h}(\tau_{2})\right]| \Omega\rangle\,.
\end{equation}
Using the expressions  (\ref{Ofh}) and (\ref{Ohf}) for the two-point functions $\langle \Omega |\mathcal{O}_{\phi}(\tau_{1})\mathcal{O}_{h}(\tau_{2})| \Omega\rangle$ and $\langle \Omega |\mathcal{O}_{h}(\tau_{1})\mathcal{O}_{\phi}(\tau_{2})| \Omega\rangle$ and commutating, we find
\begin{equation}
\frac{1}{i}\left(\langle \Omega |\mathcal{O}_{\phi}(\tau_{1})\mathcal{O}_{h}(\tau_{2})| \Omega\rangle -\langle \Omega |\mathcal{O}_{h}(\tau_{1})\mathcal{O}_{\phi}(\tau_{2})| \Omega\rangle\right)=\sum_{k} \frac{2\sin^{2}k}{\sqrt{1-2 h \cos k+h^2}}e^{-2\Lambda_{k}(\tau_{2}-\tau_{1})}\,.
\end{equation}
Substituting this result in (\ref{FfLLag}) and integrating we find
\begin{equation}
F_{\phi h}=\frac{1}{2}\sum_{k}\frac{\sin^{2}k}{\left[1-2 h \cos k+h^2\right]^{3/2}}\,,
\end{equation}
which is the exact same result as in (\ref{FPhL}).

In this manner we see again that the Lagrangian formalism yields the same results for the quantum information metric and the Berry curvature as the ones obtained by direct computation.

\section{Conclusions}

In this paper we have taken a path-integral approach for computing the quantum information metric of a quantum field theory originally developed in the context of the gauge/gravity duality, and generalized it in order to be able to compute the Berry curvature as well in a unified manner. More concretely, in the generalization presented in this paper we are able to first obtain a general expression for the quantum geometric tensor, and from this we are able to compute the quantum information metric and Berry curvature of the system. This is done for any number of dimensions of the parameter-space and for the general case of non-zero expectation values for the deformation operators. We then tested our results by applying our method to a wide array of different quantum mechanical systems where the quantum information metric and the Berry curvature can be computed exactly. In all of these cases we have observed that the Lagrangian formalism reproduces exactly the results already known in the literature. 

We can see that the advantages of this generalized approach are manifold. It allows us to compute both the quantum information metric and the Berry curvature starting from a single object, the quantum geometric tensor, in a unified manner. Moreover, since this is accomplished in a path-integral approach, it can also be applied in a more natural and straightforward manner in a quantum field theory setting. We should also mention that, from an operational point of view, this procedure allows to compute the information metric and the Berry curvature from expectation values and two-point functions with respect to the unperturbed original system, and could therefore be applied even in the case of systems where the exact solution is not known.

Looking for future lines of research, it could be interesting to see how the formalism presented in this paper could be used to study quantum phase transitions in many-body systems, given the recent interest for using the quantum fidelity and information metric as a probe for such phases. In this line of thought, it could also be of interest to see how our formalism could be applied to the study of related physical quantities in quantum information theory, such as the Fisher information metric, the Bures metric or the Loschmidt echo. Similarly, it could be interesting to formally connect this path-integral formalism with the standard Hamiltonian approach that is used in most of the literature. Other possible line of research could be to study how this Lagrangian formulation can accommodate for the study of systems with non-abelian Berry connections. Additionally, it could be very interesting to see how the unified geometric point of view of our formalism could allow for an analysis of the Berry curvature from a topological point of view which could connect with the topological classification of geometric phases studied in \cite{Kiritsis}.\footnote{We wish to thank Prof. E. Kiritsis for pointing us in this direction.}  It also could be very interesting to see how this Lagrangian approach could be applied in the study of the quantum information and the Berry phase in quantum field theories. Finally, it could be very interesting to see how the generalization presented in this paper could be applied back in the gauge/gravity duality in order to study geometric phases from an holographic point of view. All these questions call for further research. 

\section {Acknowledgments}
The authors acknowledge partial support from DGAPA-UNAM grant IN 103716 and CONACYT project 237503.
A. D. is supported by CONACyT postdoctoral grant 202954 at  IFT-UAM/CSIC and would like to thank Alberto Guijosa, Antonio Garcia-Zenteno and David Vergara of ICN-UNAM for their hospitality. 
\begin{appendix}

\section{Appendix: Quantum Information Metric and Berry Curvature}

In this section we review very briefly some facts concerning the standard derivation of the quantum geometric tensor, the quantum information metric and the Berry curvature of a quantum system. For further details the reader is referred to, e.g. \cite{Provost, Anandan, Gu, Chru}

Let us consider a quantum system in a state $|\psi\rangle$. Furthermore, let us assume that the system's Hamiltonian depends explicitly on a number of real parameters $\lambda^{a}$, $a = 1,\,\ldots N$, and the state of the system inherits this dependence and can be expressed as $|\psi (\lambda)\rangle$. Let us now consider that the system undergoes a small change in some or all of the parameters in the form $\lambda+\delta \lambda$, so that the original state $|\psi(\lambda)\rangle$ changes to $|\psi(\lambda+\delta\lambda)\rangle$. We will now focus on computing the fidelity $\mathcal{F}$ of the system when it undergoes such a change in its parameters. The $\mathcal{F}$ is defined as
\begin{equation}
\label{Fid}
\mathcal{F}(\lambda,\,\lambda+\delta\lambda)= |\langle \psi(\lambda+\delta
\lambda)|\psi(\lambda)\rangle|\,,
\end{equation}
an physically it represents a measure of the change of the state under changes in its parameter space.

 If we now make a Taylor expansion of $|\psi(\lambda+\delta\lambda)\rangle$ we obtain
\begin{equation}
|\psi(\lambda+\delta\lambda)\rangle = |\psi(\lambda)\rangle + |\partial_{a}\psi(\lambda)\rangle \delta \lambda^{a}+\frac{1}{2}|\partial_{a}\partial_{b}\psi(\lambda)\rangle \delta\lambda^{a}\delta\lambda^{b}+\cdots\,,
\end{equation}
so that
\begin{equation}
\langle \psi(\lambda)|\psi(\lambda+\delta\lambda)\rangle = 1 + \langle \psi(\lambda)|\partial_{a}\psi(\lambda)\rangle \delta\lambda^{a}+\frac{1}{2}\langle \psi(\lambda)|\partial_{a}\partial_{b}\psi(\lambda)\rangle \delta\lambda^{a}\delta\lambda^{b}+\cdots\,.
\end{equation}

If we substitute these results in the expression for the fidelity (\ref{Fid}),  we obtain up to second order
\begin{equation}
\mathcal{F}(\lambda,\,\lambda+\delta\lambda)= 1 + \frac{1}{2}G_{ab}\,\delta\lambda^{a}\delta\lambda^{b}+\cdots\,,
\end{equation}
where $G_{ab}$ is the complex \textit{quantum geometric tensor}, given by
\begin{equation}
\label{AppQGT}
G_{a b}=\langle \partial_a \psi | \partial_b \psi \rangle-\langle \partial_a \psi| \psi\rangle\langle \psi |\partial_b \psi\rangle\,.
\end{equation}

The quantum geometric tensor (\ref{AppQGT}) can be divided in its real an imaginary part. Thus, taking its real part we obtain the \textit{quantum information metric} $g_{ab}$, given explicitly as
\begin{equation}
\label{standardG}
g_{ab}\equiv\textbf{Re}\, G_{a b}= \frac{1}{2}\left(\langle \partial_a \psi |\partial_b \psi\rangle +\langle \partial_b \psi |\partial_a \psi \rangle\right)-\langle\partial_{a}\psi | \psi \rangle \langle \psi |\partial_{b}\psi\rangle\,,
\end{equation}
and which can be used to measure distances along paths in parameter space. On the other hand, the imaginary part of $G_{ab}$ will yield the \textit{Berry curvature} $F_{ab}$, given by
\begin{equation}
\label{standardF}
\frac{1}{2}F_{ab}\equiv \textbf{Im}\, G_{a b}=\frac{1}{2i}\left(\langle \partial_{a}\psi | \partial_{b}\psi\rangle -\langle \partial_{b}\psi |\partial_{a}\psi \rangle\right)\,,
\end{equation}
where the one-half factor is added by convention. 

Let us now consider the purely imaginary quantity $\langle \psi | \partial_b \psi \rangle$. We now define the \textit{Berry connection} $A_{b}$ as 
\begin{equation}
\label{connect}
A_{b}=-i\,\langle \psi | \partial_b \psi \rangle\,.
\end{equation}
Furthermore, it is easy to show that Berry curvature can be written as $F_{ab}=\partial_a A_b -\partial_b A_a$. The Berry connection is directly related to the \textit{Berry phase} $\gamma(\mathcal{C})$ of the system, given as
\begin{equation}
\gamma(\mathcal{C})=\oint_{\mathcal{C}} A\,,
\end{equation}
where we have defined the one-form $A = A_{b}dX^{b}$ and $\mathcal{C}$ is a closed curve in parameter space. The Berry phase $\gamma(\mathcal{C})$ describes a cyclic adiabatic evolution along $\mathcal{C}$. By making use of Stokes theorem, we can rewrite the Berry phase as
\begin{equation}
\gamma(\mathcal{C})=\int_{\Sigma} F\,,
\end{equation}
where $\Sigma$ is a two-dimensional manifold such that $\partial \Sigma = \mathcal{C}$ and where we have defined the two-form
\begin{equation}
\label{2F}
F = \frac{1}{2}F_{ab}\,dX^{a}\wedge dX^{b}\,.
\end{equation}
\end{appendix}

\end{document}